\documentclass[final,3.5p,times,preprint]{elsarticle}

\usepackage{amssymb,amsmath,float}
\usepackage{lipsum}
\usepackage{cancel}
\usepackage[T1]{fontenc}

\usepackage{tcolorbox}
\newtcolorbox{mybox}[1]{
		colback=black!5!white, 
		colframe=black, 
		fonttitle=\bfseries,title=#1}

\usepackage{hyperref}

\newcommand{\be}{\begin{eqnarray}}
\newcommand{\ee}{\end{eqnarray}}
\newcommand{\ba}{\begin{array}}
\newcommand{\ea}{\end{array}}
\newcommand{\pa}[1]{\left(#1\right)}

\newcommand{\ds}{\displaystyle}

\usepackage{listings}
\usepackage{xcolor}

\definecolor{codegreen}{rgb}{0,0.6,0}
\definecolor{codegray}{rgb}{0.5,0.5,0.5}
\definecolor{codepurple}{rgb}{0.58,0,0.82}
\definecolor{backcolour}{rgb}{0.95,0.95,0.92}

\lstdefinestyle{mystyle}{
    backgroundcolor=\color{backcolour},   
    commentstyle=\color{codegreen},
    keywordstyle=\color{magenta},
    numberstyle=\tiny\color{codegray},
    stringstyle=\color{codepurple},
    basicstyle=\ttfamily\footnotesize,
    breakatwhitespace=false,         
    breaklines=true,                 
    captionpos=b,                    
    keepspaces=true,                 
    numbers=left,                    
    numbersep=5pt,                  
    showspaces=false,                
    showstringspaces=false,
    showtabs=false,                  
    tabsize=2,
    inputencoding = utf8,  
    extendedchars = true,  
    literate      =        
      {á}{{\'a}}1  {é}{{\'e}}1  {í}{{\'i}}1 {ó}{{\'o}}1  {ú}{{\'u}}1
      {Á}{{\'A}}1  {É}{{\'E}}1  {Í}{{\'I}}1 {Ó}{{\'O}}1  {Ú}{{\'U}}1
      {à}{{\`a}}1  {è}{{\`e}}1  {ì}{{\`i}}1 {ò}{{\`o}}1  {ù}{{\`u}}1
      {À}{{\`A}}1  {È}{{\'E}}1  {Ì}{{\`I}}1 {Ò}{{\`O}}1  {Ù}{{\`U}}1
      {ä}{{\"a}}1  {ë}{{\"e}}1  {ï}{{\"i}}1 {ö}{{\"o}}1  {ü}{{\"u}}1
      {Ä}{{\"A}}1  {Ë}{{\"E}}1  {Ï}{{\"I}}1 {Ö}{{\"O}}1  {Ü}{{\"U}}1
      {â}{{\^a}}1  {ê}{{\^e}}1  {î}{{\^i}}1 {ô}{{\^o}}1  {û}{{\^u}}1
      {Â}{{\^A}}1  {Ê}{{\^E}}1  {Î}{{\^I}}1 {Ô}{{\^O}}1  {Û}{{\^U}}1
      {œ}{{\oe}}1  {Œ}{{\OE}}1  {æ}{{\ae}}1 {Æ}{{\AE}}1  {ß}{{\ss}}1
      {ç}{{\c c}}1 {Ç}{{\c C}}1 {ø}{{\o}}1  {Ø}{{\O}}1   {å}{{\r a}}1
      {Å}{{\r A}}1 {ã}{{\~a}}1  {õ}{{\~o}}1 {Ã}{{\~A}}1  {Õ}{{\~O}}1
      {ñ}{{\~n}}1  {Ñ}{{\~N}}1  {¿}{{?`}}1  {¡}{{!`}}1
      {°}{{\textdegree}}1 {º}{{\textordmasculine}}1 {ª}{{\textordfeminine}}1
}

\begin{document}

\begin{frontmatter}

  \title{GWDALI: A Fisher-matrix based software for gravitational wave parameter-estimation beyond Gaussian approximation}

\author[ufrn]{Josiel Mendonça Soares de Souza}
\affiliation[first]{organization={Departamento de F\'\i sica Te\'orica e Experimental},
  city={Natal},
  state={RN},
  country={Brazil}}

\author[unesp]{Riccardo Sturani}
\affiliation[ufrn]{organization={Instituto de F\'\i sica Te\'orica, UNESP/ICTP-SAIFR},
  city={S\~ao Paulo},
  state={SP},
  country={Brazil}}

\begin{abstract}
  We introduce \textit{GWDALI}, a new Fisher-matrix, python based software that
  computes
  likelihood gradients to forecast parameter-estimation precision
  of arbitrary network of terrestrial gravitational wave detectors observing
  compact binary coalescences.
  The main new feature with respect to analogous software is to assess
  parameter uncertainties beyond Fisher-matrix approximation, using the
  derivative approximation for Likelihood (DALI).
  The software makes optional use of the LSC algorithm
  library {\tt LAL} and the stochastic
  sampling algorithm {\tt Bilby}, which can be used to perform
  Monte-Carlo sampling of exact or approximate likelihood functions.
  As an example we show comparison of estimated precision measurement of selected
  astrophysical parameters for both the actual likelihood, and for a variety of
  its derivative approximations, which turn out particularly useful when the
  Fisher matrix is not invertible.
\end{abstract}



\end{frontmatter}

\tableofcontents

\section{Introduction}
\label{sec:intro}

The advent of Gravitational Wave (GW) astronomy \cite{LIGOScientific:2021djp}, with the
detections made by the LIGO \cite{TheLIGOScientific:2014jea} and Virgo \cite{TheVirgo:2014hva} detectors, naturally calls for investigating which physical quantity and how well
it can be measured with the observations made possible in this new gravitational channel.

The detector network should be enriched by the Japanese KAGRA
\cite{KAGRA:2020tym}
already in the ongoing fourth science run of the so-called \emph{second generation}
GW detectors, and by the Indian Indigo \cite{indigo}
by the end of the current decade.
For the next decade, additional third generation detectors are foreseen, namely
the European Einstein Telescope (ET) \cite{Maggiore:2019uih}
and the North-American Cosmic Explorer (CE) \cite{Hall:2022dik}.

While detections in the future may not be limited to compact binary coalescences, they represent the totality of detections so far, and they
are expected to be the key observables for future observatories as well.
It is then natural to ask ourselves for expectations and forecast of precision measures
of astrophysical source parameters, how they depends on the intrinsic
detector feature, like
spectral noise density, and localization and orientation of the detectors
composing the network.

Parameter estimation is naturally performed within a Bayesian inference
framework, which allows determination of most likely values of signal
model parameter as well as their confidence interval.
A common tool for rapid precision forecast is the \emph{Fisher matrix}
approximation, which consists in approximating the logarithm of the likelihood
in the proximity of its maximum value with a Taylor expansion truncated
at quadratic order.
By inverting the matrix of second derivatives of the log-likelihood with respect
to all parameters one can then obtain the \emph{covariance} matrix, which
gives direct assessment of marginalized uncertainties of physical parameters
\cite{Trotta:2008qt}.
On one hand the Fisher matrix approximation is ubiquitous in data analyses,
as an invaluable tool to provide analytic and easy-to-compute
estimates of parameter uncertainties, on the other hand there are cases in which it can fail,
e.g. in presence of approximately flat directions and/or correlations among parameters,
giving rise to zero determinant second derivative matrix.

In this case it is natural to investigate how higher order approximations
may perform with respect to the quadratic one and/or the
complete likelihood. The goal of the present work is to introduce
a tool to perform a suitable Taylor expansion of the derivative approximation
of the log-likelihood. This is implemented in the publicly available python code GWDALI\footnote{\hyperlink{github.com/jmsdsouzaPhD/GWDALI/}{github.com/jmsdsouzaPhD/GWDALI/}}, complementing
standard Fisher matrix codes already available in literature like \textit{GWBENCH} \cite{Borhanian:2020ypi},
\textit{GWFISH} \cite{Dupletsa:2022scg}, and \textit{GWFAST} \cite{Iacovelli:2022mbg}.

In particular we borrow here the idea to use higher (than second) derivative approximations to the log-likelihood from \cite{Sellentin:2014zta}, where the Derivative
Approximation for LIkelihood (DALI) was first proposed
in a general context and applied to cosmological examples,
and from 
\cite{Wang:2022kia}, where a first example of DALI applied to GW source parameter inference
has been proposed.

The paper is structured as follows: in Sec. \ref{sec:problems} we give an
overview of parameter estimation for GW signal from coalescing binaries highlighting some known limit of the Fisher-matrix approximation, followed in Sec.~\ref{sec:dali} by a description of the DALI method.
In Sec.~\ref{sec:python} we give an example of the benefits of using GWDALI for estimating parameter uncertanties and finally in Sec.~\ref{sec:concl} we outline possible
future applications and development of the GWDALI software.

\section{Astrophysical parameter estimation for coalescing binaries and Fisher matrix}
\label{sec:problems}

The coalescence of compact binary systems can be divided in three phases:
inspiral, merger and ring-down.
In particular the inspiral phase admits an analytic description in terms of approximations to General Relativity, the most successful of which in building waveform
models has been the post-Newtonian approximation \cite{Blanchet:2013haa,Isoyama:2020lls}.
The subsequent merger phase is in principle determined by the inspiral one,
of which it is the continuation, and it has been described solved with numerical
methods \cite{Boyle:2019kee}.
Finally the ring-down phase also admits an analytic description, in terms of
damped exponentials, and several phenomenological models exist nowadays providing
complete analytic waveforms encompassing the three phases, see e.g.~\cite{Bohe:2016gbl,Khan:2015jqa}. 
An exhaustive set of waveform approximant is available in the LSC algorithm library ({\tt LAL})
\cite{lalsuite}, which are callable by python or C codes.

The binary systems of interest for current detectors are expected to be
in circular orbits, as angular momentum is more efficiently radiated
than energy. Orbits are indeed expected to circularize in the early stage
of inspiral \cite{Peters:1963ux}, leaving little or no eccentricity when reaching the LIGO/Virgo
band ($10-10^4$ Hz), which is not too distant from the merger frequency $f_m$,
which can be roughly approximated by $(2\pi M)^{-1}$, being $M$ the total mass of the binary system, obtaining
$f_m\sim 32$kHz$(M/M_\odot)^{-1}$.\footnote{We use ``geometric units'', with $G=1=c$.}
The astrophysical parameters defining a circular binary coalescences
are 15 \cite{Finn:1992xs}, organized in two categories \cite{Owen:1995tm}
according to if they affect the morphology of the waveform
(intrinsic), or just its overall shape (extrinsic).
The 15 parameters are summarized in
Tab.~\ref{tab:15pars}.

\begin{table}[t]
  \begin{center}
  \begin{tabular}{||c|c||}
    \hline
    Intrinsic parameters & Extrinsic parameters\\
    \hline
    $M_c$, $\eta$, $\vec S_1$, $\vec S_2$ &
    $d_L$, $\psi$, $\iota$, $\phi$, $\alpha$, $\beta$, $t$\\
    \hline
  \end{tabular}    
  \caption{Parameters defining the observation of a binary system observation
    (whose constituents
    are treated as point-like object),  divided between intrinsic ($m_{1,2}$
    are the constituent masses, $\vec S_{1,2}$
    are the binary constituent spin vectors) and extrinsic (see text).}
  \label{tab:15pars}
  \end{center}
\end{table}

\begin{figure}
  \centering
  \includegraphics[width=.6\linewidth]{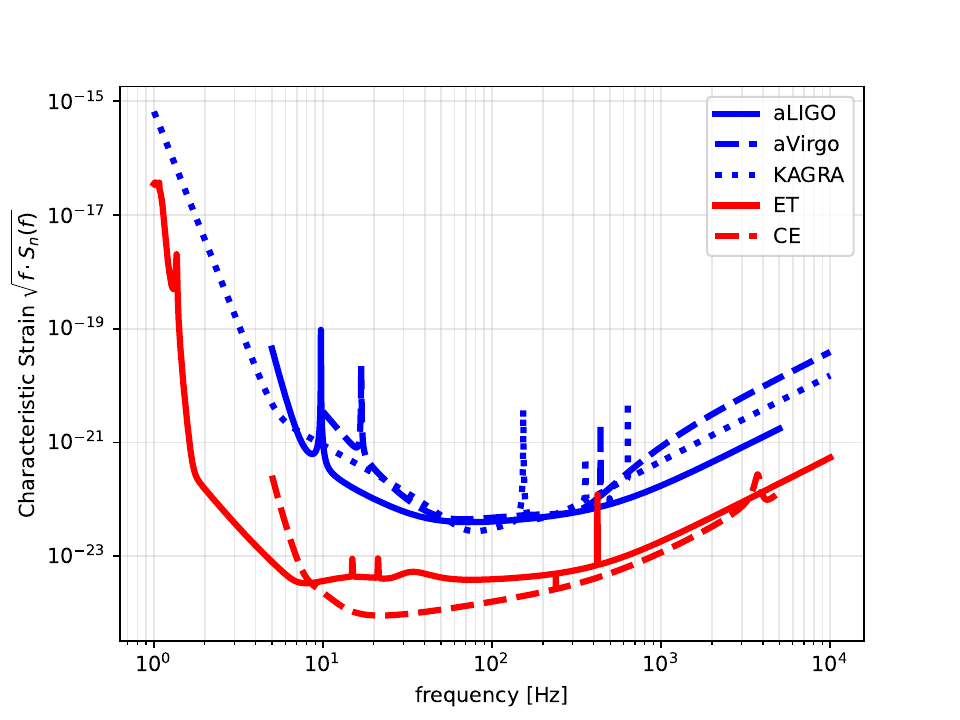}
  \caption{ET \cite{etd} and CE \cite{Srivastava:2022slt} noise spectral densities.}
  \label{fig:sns}
\end{figure}

Given the time series of the output of the $i$-th detector $d_i$ and a
\emph{template} signal $h_{t_i}$, the log-likelihood for astrophysical parameter
inference is proportional the norm-squared $||d_i-h_{t_i}||^2$, where the norm is inherited
from the scalar product
\begin{equation}
  \label{eq:mf}
\langle h_1,h_2\rangle=2\int_0^\infty \frac{\tilde h_1(f) \tilde h_2^*(f)+\tilde h_1^*(f) \tilde h_2(f)}{S_n(f)}df\,,
\end{equation}
where $S_n(f)$ is the one sided \emph{spectral noise density} of the detector defined
by averaging the noisy detector output:
\be
\langle \tilde n(f)\tilde n^*(f')\rangle\equiv \frac 12 S_n(f)\delta(f-f')\,.
\ee
The Einstein Telescope (ET) \cite{Maggiore:2019uih} and Cosmic Explorer (CE)
\cite{Evans:2021gyd} sensitivity curves adopted in this work are shown
in Fig.~\ref{fig:sns}.

The detector response to a GW is a linear combination of the two polarizations with weights given by the \emph{pattern functions} $F_{+,\times}$, which depend
on sky localization of the source (defined by polar angles $\alpha,\beta$) and
the polarization angle $\psi$:
\begin{equation}
  h_{d_i}=F_+(\alpha,\beta,\psi)h_++F_\times(\alpha,\beta,\psi)h_\times\,.
\end{equation}

The specific form of the pattern functions depend on the type of detector
considered. For interferometric detectors with opening angle $\Omega$
they are, see ch. 7 of \cite{Maggiore:2007ulw},
\be
\ba{rcl}
F_+&=&f_+\cos(2\psi)+f_\times\sin(2\psi)\,,\\
F_\times&=&f_\times\cos(2\psi)-f_+\sin(2\psi)\,,
\ea
\ee
with
\be
\ba{rcl}
f_+&\equiv&\sin\Omega\frac 12(1+\cos^2\beta)\cos(2\alpha)\,,\\
f_{\times}&\equiv&-\sin\Omega\cos\beta\sin(2\alpha)\,.
\ea
\ee
In the lowest order approximation (quadrupole formula) the polarizations for
the inspiral phase admit a simple analytic form, see e.g. ch.4 of ~\cite{Maggiore:2007ulw}
\be
\label{eq:wfs}
\ba{rcl}
\ds\tilde h_+(f)&=&\ds\frac{1+\cos^2\iota}2\pi^{-2/3}\pa{\frac 5{24}}^{1/2}
\frac{\mathcal{M}_c^{5/6}f^{-7/6}}{d_L}e^{i\Phi(f)}\,,\\ 
\ds\tilde h_\times(f)&=&\ds\cos\iota\,\pi^{-2/3}\pa{\frac 5{24}}^{1/2}
\frac{\mathcal{M}_c^{5/6}f^{-7/6}}{d_L}i e^{i\Phi(f)}\,, 
\ea
\ee
where $\mathcal{M}_c$ is the redshifted \emph{chirp mass} defined by
$\mathcal{M}_c\equiv \eta^{3/5}M(1+z)$, with $M$ the \emph{intrinsic} total mass $M\equiv m_1+m_2$,
$\eta$ the symmetric mass ratio $\eta\equiv m_1m_2/M^2$, $z$ is the redshift
and $\Phi(f)$ has an expansion known up to 4th perturbative order in terms of
the PN expansion parameter $v^2\equiv (\pi Mf)^{2/3}$, see e.g.~\cite{Blanchet:2023sbv} and depends on $\eta$ and linearly on the arrival time $t$ and the phase $\phi$.
Note that $\iota,\phi,\psi$ compose the three Euler angles defining the orientation of the source axis triad with respect to the observation one \cite{deSouza:2023gjv}.

The log-likelihood vanishes when the template matches exactly the data, the
Fisher matrix for the $i$-th detector being
\be
{\cal F}_{i,ab}=4{\cal R}e\int_0^{\infty}\frac{\partial_{\theta_a}\tilde h_{t_i}
  \partial_{\theta_b}\tilde h_{t_i}^*}{S_n(f)} df\,,
\ee
where dependence of $h_{t_i}$ (and $\tilde h_{t_i}$) upon parameters $\theta$ is understood in the notation.
The $k$-th element of the diagonal of the inverse of the Fisher matrix, the \emph{covariance} matrix, returns
marginalized uncertainties for the $k$-th
variable, which then crucially depend on correlation among parameters.

\begin{figure}[H]
    \centering
    \includegraphics[width=.49\linewidth]{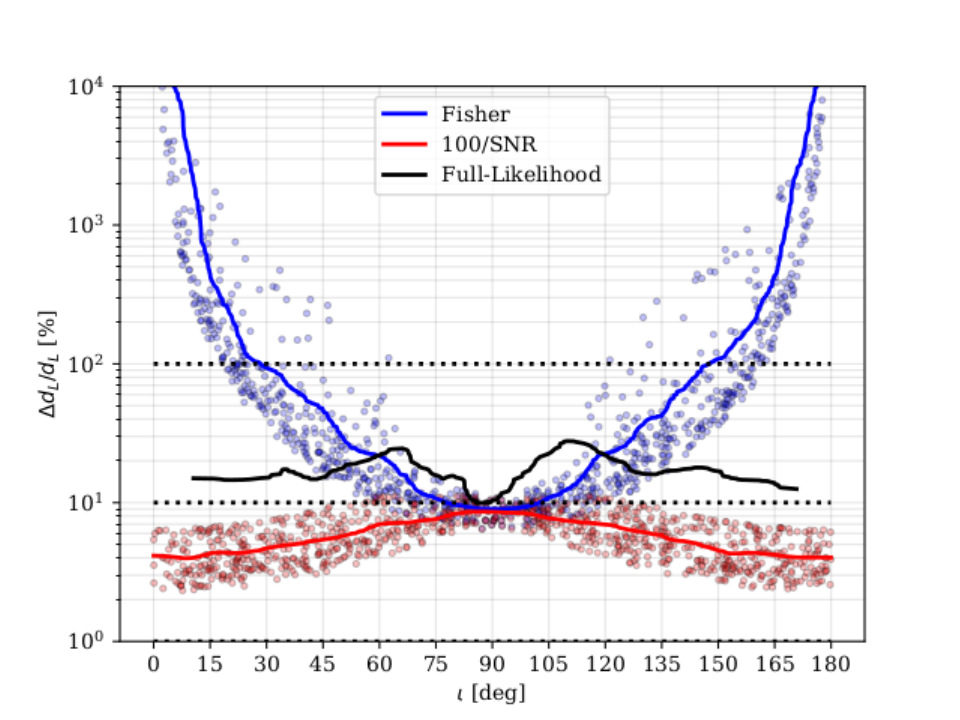}
    \includegraphics[width=.49\linewidth]{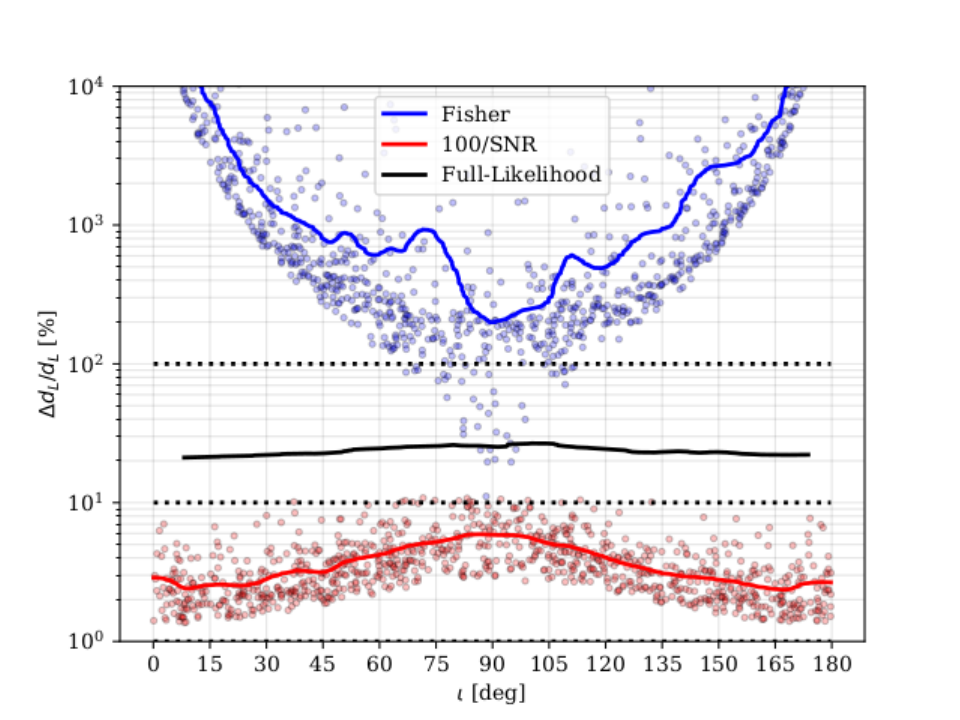}
    \caption{Example of $\Delta d_L/d_L$ vs. $\iota$ for a 4-dimensional Fisher matrix in the
      parameters: $\{d_L, \iota, \psi, \phi\}$ from
      GWFISH code (blue line) \cite{Dupletsa:2022scg}, full likelihood for the same parameters (black) and for the analytic curve given by $100/SNR$ (red). GW signals are generated by equal mass
      spin-less binaries with $M=3M_\odot$ with TaylorF2 approximant \cite{Sathyaprakash:2009xs}, averaged over sky position and
      inclination angles for ET (left) and CE (right).}
    \label{fig:gwfish}
\end{figure}

One of the obvious shortcomings of the Fisher matrix approximation is encountered
when the Fisher matrix itself is not invertible, hence it is not possible to find the covariance
matrix.
This is exemplified in Fig.~\ref{fig:gwfish}, where the result for $d_L$ of
the four dimensional covariance matrix in $\{d_L,\iota,\psi,\phi\}$ are displayed,
as obtained with the code GWFISH for the ET detector (made
of three interferometers arranged in a equilater triangles) and for CE (one $L$-shaped interferometer).

The increase of $d_L$ uncertainty, which is unbound for $\iota\to 0,\pi$ in
the Fisher matrix approximation, is due to $det({\cal F})\to 0$ as $\iota \to 0,\pi$,
signaling a degenerate direction in the parameter space, and invalidating
the Gaussian approximation the Fisher matrix approach is based on.
The fact that this is a spurious behaviour due to the Fisher approximation of the exact likelihood, is highlighted by the analaog quantity
obtained by MonteCarlo sampling of the likelihood, black
line in Fig.~\ref{fig:gwfish}. 
We discuss in Sec.~\ref{sec:dali} a proposal about how to handle this problem.

\subsection{Analytic Fisher Matrix Inversion}
\label{ssec:inversion}

The inverse of the Fisher matrix is the covariance matrix, whose diagonal
entries give the fully marginalized 1$\sigma$ errors (in the Gaussian approximation) of the corresponding
parameters, which is probably the most useful property of the Fisher method.
The diagonal entries of the covariance matrix are different depending
on how many parameter one keeps fixed to the maximum likelihood values, i.e.~they depend on the dimensionality of the Fisher matrix.
As an example, the $d_L$ 1$\sigma$-uncertainty with all other parameters fixed
can be computed as:
\begin{equation}
  \left(\frac{\Delta d_L}{d_L}\right)^2 = \frac{1}{SNR^2} =
  \frac{4||k||^{-2}}{\left[(1+\cos^2\iota)^2 F_+^2 + 4cos^2\iota F_{\times}^2\right]}\,,
\end{equation}
where using the inspiral expressions (\ref{eq:wfs}) one has
\begin{equation}
  ||k||^2 \equiv \frac{5}{6}\frac{(G\mathcal{M}_c)^{5/3}}{\pi^{4/3}d_L^2}\int_0^{\infty} \frac{f^{-7/3}}{S_n(f)}df\,,
\end{equation}
and the $SNR$ is defined as the norm of the GW signal $h$ given the scalar product in eq.~(\ref{eq:mf}).

Marginalizing over $\iota$, instead of pinning it to its maximum likelihood
value, shows the effect of correlations into determining parameter uncertainty.
In this case, one can then read the $d_L$ uncertainty as
\begin{equation}
  \pa{\frac{\Delta d_L}{d_L}}^2 = \frac{Cov(d_L,d_L)}{d_L^2} = \frac{\langle \partial_{\iota} h, \partial_{\iota} h\rangle}{d_L^2det({\cal F})}
  = ||k||^{-2}\frac{4(\cos^2\iota F_+^2 + F_{\times}^2)}{\sin^4\iota(F_+F_{\times})^2}\,.
\end{equation}

\begin{figure}[H]
  \centering
  \includegraphics[width=.49\linewidth]{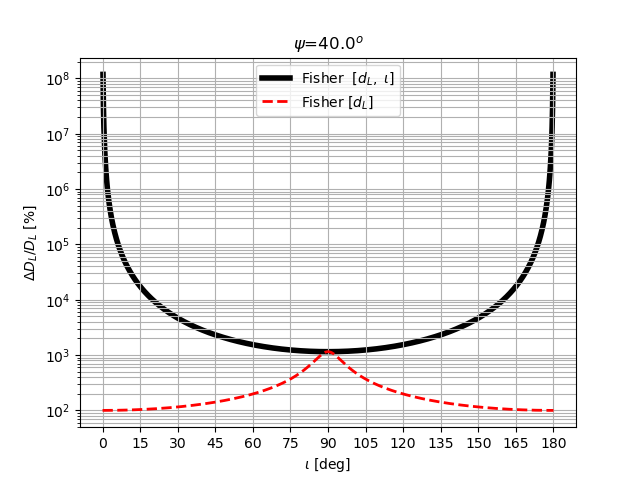}
  \includegraphics[width=.49\linewidth]{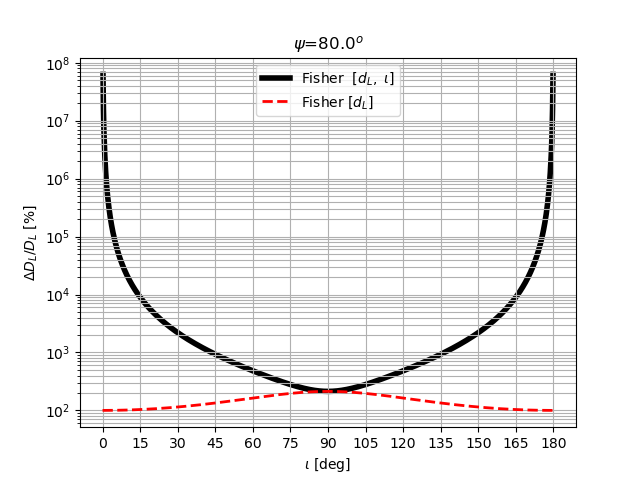}
  \caption{1$\sigma$ relative uncertainty in $d_L$ fixing all other parameters
  (dashed red) or marginalizing over $\iota$ (solid black).}
  \label{fig:corr}
\end{figure}

This is a simple example of how correlations among parameters can degrade the
precision of individual parameters. In particular
Figs.~\ref{fig:corr} report the analytic Fisher matrix prediction
of the uncertainty estimate of $d_L$ as a function of the injected value
of $\iota$ in two cases: 1-dimensional Fisher matrix and bi-dimensional
for parameters $d_L,\iota$ (sources are located above a CE-like
detector).

The presence of a zero eigenvalue in the Fisher matrix for $\iota\to 0,\pi$
denotes the existence of a flat direction, invalidating the Gaussian
approximation. In this example such breakdown manifests itself by the
divergence of the Fisher matrix prediction of $\Delta d_L$.

Note that standard recipes for obtaining the (pseudo-)inverse of a
zero-determinant Fisher matrix are adopted in e.g. GWFISH \cite{Dupletsa:2022scg}
and GWFAST \cite{Iacovelli:2022mbg}, where the Penrose method for generalized
inverse  \cite{Penrose:1955vy} is implemented.
Such method consists of excluding from the
inversion the subspace corresponding to the kernel of the Fisher matrix.
However identifying a zero eigenvalue numerically depends on an arbitrary
threshold.\footnote{In GWFISH the threshold for a "numerical zero" is set to $10^{-10}$.}
We propose in the next section the application of higher derivative likelihood
approximation to GW parameter estimation.

\section{DALI algorithm for Gravitational Waves}
\label{sec:dali}

Following the introduction of higher derivative in the likelihood expansion
in \cite{Sellentin:2014zta},
we now introduce the GWDALI implementation of likelihood beyond Gaussian order.
Let $\theta$ be the set of GW parameters listed in Tab.~\ref{tab:15pars}.
The Fisher matrix approximation relies on a Taylor expansion of the log-likelihood
around the maximum likelihood value. Given the definition (\ref{eq:mf}),
the log-likelihood as the norm-squared of the difference between data and template, one has the straightforward Taylor expansion beyond quadratic order
\be
\label{eq:taylor}
\ba{rcl}
\log\mathcal{L} & =&\ds-\frac 12||d_i-h_{t_i}(\theta)||^2\\
&\simeq&\ds \log\mathcal{L}_{0}+\cancel{\sum_{a}\left(\partial_a\log\mathcal{L}\right)_{0}\Delta\theta_a}+\frac{1}{2!}\sum_{a,b}\left(\partial_{a}\partial_{b}\log\mathcal{L}\right)_{0}\Delta\theta^{ab}\\
&&\ds +\frac{1}{3!}\sum_{a,b,c}\left(\partial_{a}\partial_{b}\partial_{c}\log\mathcal{L}\right)_{0}\Delta\theta^{abc}\\
&&\ds +\frac{1}{4!}\sum_{a,b,c,d}\left(\partial_{a}\partial_{b}\partial_{c}\partial_{d}\log\mathcal{L}\right)_{0}\Delta\theta^{abcd}\\
&&\ds+\frac{1}{5!}\sum_{a,b,c,d,e}\left(\partial_{a}\partial_{b}\partial_{c}\partial_{d}\partial_e\log\mathcal{L}\right)_{0}\Delta\theta^{abcde}+\ldots\,,
\ea
\ee

where $\Delta\theta^a\equiv\theta^a-\theta^a_0$,
$\Delta\theta^{a_1\dots a_n}\equiv \Delta \theta^{a_1}\dots\Delta\theta^{a_n}$ and the subscript ``$0$''
stand for maximum likelihood value of the parameters. The first order term of the
Taylor expansion has been slashed out as it vanishes.

One can then define derivative tensors beyond the Fisher matrix
\be
\label{eq:beyond}
\ba{ll}
\text{Fisher Matrix: }{\cal F}_{ab} &\ds\equiv -\left(\partial_a\partial_b\log\mathcal{L}\right)_{0}=4\sum_{i=1}^{N_{det}}{\cal R}e
\int_0^\infty \left(\frac{\partial \tilde h_{t_i}}{\partial\theta^a} \frac{\partial \tilde h_{t_i}}{\partial\theta^b}\right)\frac{df}{S_n(f)}\\
&\ds \equiv\left\langle \partial_ah|\partial_bh\right\rangle\,,\\
\text{Flexion Tensor: }{\cal S}_{abc} &\ds\equiv-\left(\partial_a\partial_b\partial_c\log\mathcal{L}\right)_{0}\,,\\
	\text{Quarxion Tensor: }{\cal Q}_{abcd} &\ds\equiv-\left(\partial_a\partial_b\partial_c\partial_d\log\mathcal{L}\right)_{0}\,,\\
	\text{P-Tensor: }\mathcal{P}_{abcde} &\ds \equiv-\left(\partial_a\partial_b\partial_c\partial_d\partial_e\log\mathcal{L}\right)_{0}\,,\\
	\text{H-Tensor: }\mathcal{H}_{abcdef} &\ds \equiv-\left(\partial_a\partial_b\partial_c\partial_d\partial_e\partial_f\log\mathcal{L}\right)_{0}\,.
\ea
\ee

The explicit expression of the symmetric derivative tensors at third order
Taylor expansion ($O(\Delta\theta^3)$), or {\bf Flexion} term,
is given by the addition of 3 terms that can be arranged as follows:
  \be
  {\cal S}_{abc}=\left\langle \partial_ah|\partial_b\partial_ch\right\rangle_{0}+(a\leftrightarrow b)+(a\leftrightarrow c)\,,
  \ee
  and its contribution to the likelihood is given by
  \be
  \ba{cl}
    \log\mathcal{L}_3 &\ds \equiv -\frac{1}{6}\sum_{a,b,c}{\cal S}_{abc}\Delta\theta^{abc}\\
&\ds =-\frac{1}{2}\sum_{a,b,c}\left\langle \partial_ah|\partial_b\partial_ch\right\rangle_{0}\Delta\theta^{abc}\,,
    \ea
    \ee
where at most second derivative of the waveform are involved.
Note that an approximate log-likelihood truncated at $O(\Delta\theta^3)$
is not negative semi-definite, causing a fundamental problem in the probabilistic
interpretation of the likelihood.

However, as explained in \cite{Sellentin:2014zta}, this problem is fixed by
truncating the expansion
in number of derivatives, rather than in powers of $\Delta\theta$.
Going beyond Gaussian level then requires to collect all terms involving
a given number of derivatives.
To have a consistent expansion with two derivatives one needs to include
the relevant part of the {\bf Quarxion} term:
\be
\ba{cl}
{\cal Q}_{abcd} =&\ds \left[\left\langle \partial_a\partial_bh|\partial_c\partial_dh\right\rangle
        +(b\leftrightarrow c)+(b\leftrightarrow d)\right]\\
&+\left[\left\langle \partial_ah|\partial_b\partial_c\partial_dh\right\rangle _{0}+(a\leftrightarrow b)+(a\leftrightarrow c)+(a\leftrightarrow d)
          \right]\,,
\ea
\ee
where the terms in the last line involve the maximum number of derivatives (3)
at this $\Delta\theta$-expansion order (4).

The Quarxion contribution to the likelihood is
\be
\label{eq:quarx}
\ba{cl}
\log\mathcal{L}_4 &\ds \equiv-\frac 1{24}\sum_{abcd}{\cal Q}_{abcd}\Delta\theta^{abcd}\\
	&\ds =-\sum_{a,b,c,d}\left[\frac{1}{8}
  \left\langle \partial_a\partial_bh|\partial_c\partial_dh\right\rangle_0
  +\frac{1}{6}\left\langle \partial_ah|\partial_b\partial_c\partial_dh\right\rangle_0\right] \Delta\theta^{abcd}\,.
\ea
\ee
Only the first term of the square bracket in the second line of (\ref{eq:quarx})
contributes to the two derivative expansion term, while the remaining one
contributes to the three derivative expansion terms, which is treated in
\ref{app:PH}.

\subsection{Approximate likelihood: Doublet-DALI}

The \textbf{Doublet-DALI} is obtained from Taylor expansion of
the log-likelihood and collecting terms involving up to 2nd-order derivatives,
hence it is obtained by adding the Flexion term and the two-derivative part of
the Quarxion term. The \textbf{Triplet-DALI} is obtained by including terms
with up to 3rd-order derivatives, which involve parts of the Quarxion, the P-
and the H-Tensor, which are treated in \ref{app:PH}.

The complete result for the Doublet-DALI is
\be
\ba{rl}
  \ds\log\mathcal{L}_{Doublet} =&\ds \log\mathcal{L}_{0}
  -\frac{1}{2}\sum_{a,b}\left\langle \partial_ah|\partial_bh\right\rangle_{0}\Delta\theta^{ab}\nonumber \\
  &\ds -\frac{1}{2}\sum_{a,b,c}\left\langle\partial_ah|\partial_b\partial_ch\right\rangle_{0}\Delta\theta^{abc}
  -\frac{1}{8}\sum_{a,b,c,d}\left\langle \partial_a\partial_bh|\partial_c\partial_dh\right\rangle _{0}\Delta\theta^{abcd}\,.
\ea
\ee

\section{GWDALI python package}
\label{sec:python}

The main call to the GWDALI package can be exemplified as follows:

\begin{lstlisting}[language=Python]
Result = GWDALI(Detection_Dict, 
                FreeParams, 
                detectors, 
                approximant = 'TaylorF2',
                fmin  = 1., 
                fmax  = 1.e4, 
                fsize = 3000, 
                dali_method    = 'Doublet',
                sampler_method = 'nestle', 
                npoints      = 100,
                save_samples = False, 
                save_cov     = False, 
                save_fisher  = False,
                plot_corner  = False,
                hide_info    = False,
                rcond        = 1.e-4,
                index		 = 1)
                
# np: length of FreeParams
# Result['Fisher']     : Fisher Matrix >> shape: (np,np)
# Result['CovFisher']  : Covariance from Fisher (inversion) >> shape: (np,np)
# Result['Covariance'] : Covariance from DALI >> shape: (np,np)
# Result['Samples']    : samples >> np columns
# Result['Error']      : list of 1-sigma uncertainties of parameters
# Result['Recovery']   : list of recovered params              
\end{lstlisting}

The \verb|detection_dict| input parameter is a dictionary containing all injection parameter
names and values. The \verb|detectors| input contains all information about detector (location on earth, opening angle between arms, spectral noise density).
The user defines via the dictionary \verb|FreeParams| the variables
for which marginalized uncertainties have to be computed.
In the case of a beyond Fisher matrix approximation of the likelihood,
i.e. Doublet or Triplet, a MonteCarlo sampling is run to estimate the
uncertainties and the maximum likelihood estimators of the free parameters.
A Bayesian inference code with sampler method specified in
\verb|sampler_method| is run via {\tt Bilby} \cite{Ashton:2018jfp} to
perform a search over the parameters specified in \verb|FreeParams|.

Some features of the GWDALI software that may be worth emphasising are:
\begin{itemize}
    \item The user can choose among the methods: Fisher, Fisher\_Sampling , Doublet or Triplet for the Likelihood derivative approximation, the last three involving a
    stochastic sampler to compute the (marginalized) likelihood, while the first one is analytic in the numerical expression of the waveform derivatives.
    \item User defined detector location coordinate, orientation and geometry.
    \item The GW waveform can be generated via a {\tt LAL} call, hence
      all approximants and features available in {\tt LAL} can be used .
    \item The code returns MonteCarlo samples, Fisher and covariance
      matrix, as well as the SNR.
\end{itemize}

\begin{figure}[H]
    \centering
    \includegraphics[width=.32\linewidth]{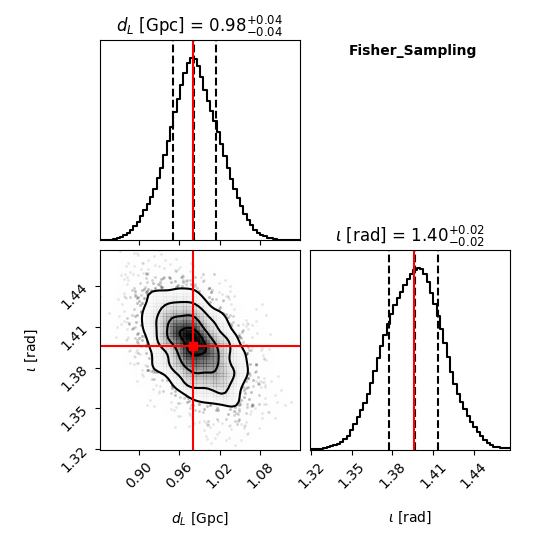}
    \includegraphics[width=.32\linewidth]{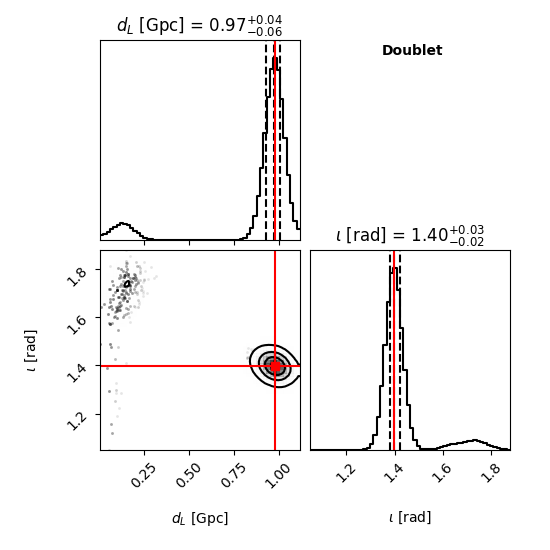}
    \includegraphics[width=.32\linewidth]{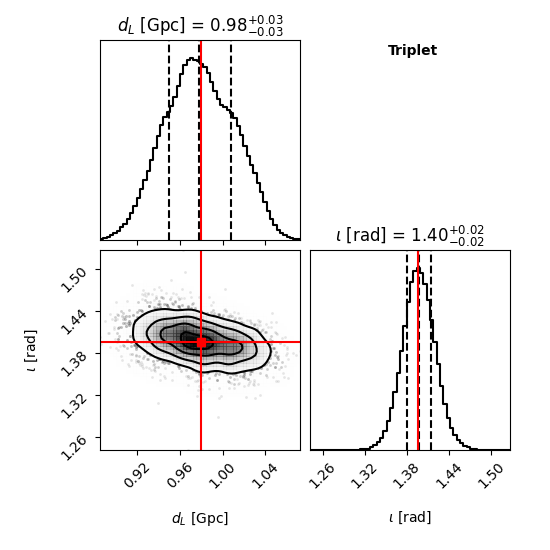}
    \caption{Example of Bayesian inference results obtained with one Einstein Telescope (triangle-shaped) detector, searching over parameters $\{d_L,\iota\}$,
      obtained with GWDALI.}
    \label{fig:Bayes_ET}
\end{figure}

\begin{figure}[H]
    \centering
    \includegraphics[width=.32\linewidth]{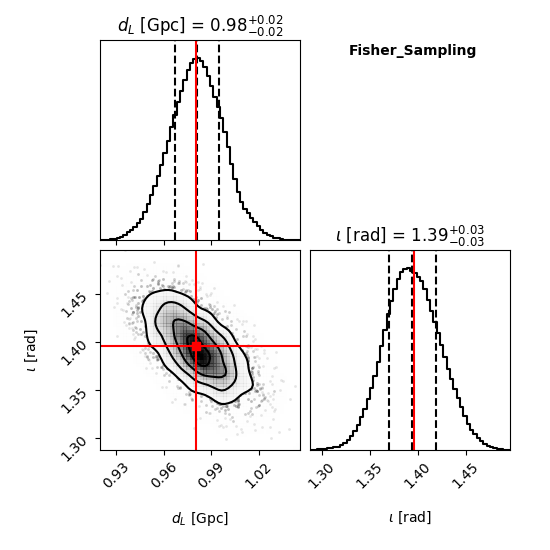}
    \includegraphics[width=.32\linewidth]{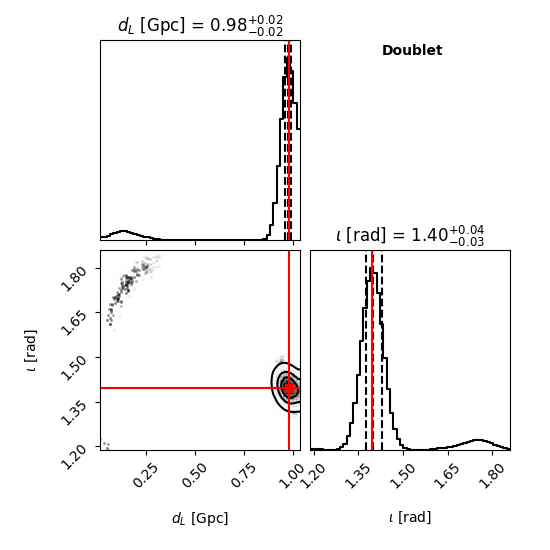}
    \includegraphics[width=.32\linewidth]{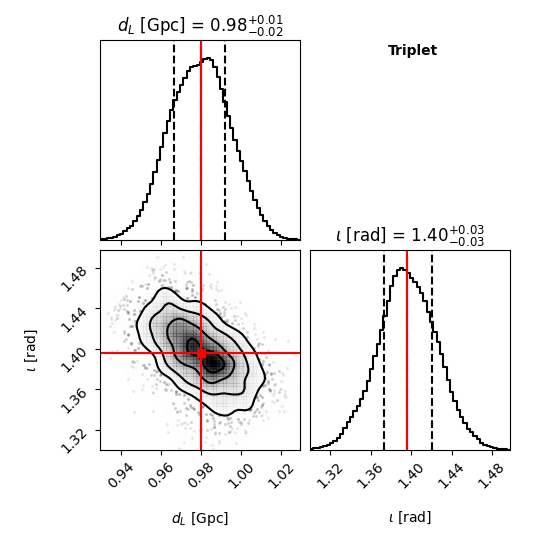}
    \caption{Example of Bayesian inference results obtained with one Cosmic Explorer ($L$-shaped) detector, searching over parameters $\{d_L,\iota\}$,
      obtained with GWDALI.}
    \label{fig:Bayes_CE}
\end{figure}

In Figs.\ref{fig:Bayes_ET},\ref{fig:Bayes_CE} we report examples of
2-dimensional posteriors for the parameter pair $\{d_L,\iota\}$ for the two detectors whose spectral noise density are reported in Fig.\ref{fig:sns}.

\subsection{Example of luminosity distance uncertainty}
\label{ssec:exs}

As an example, we run GWDALI with 3 free parameters $\{d_L, \iota, \psi\}$ and we compare the result of the Fisher matrix approximation with those of higher derivative
approximations of the likelihood and with the outcome of a Bayesian inference
code run on the exact likelihood.
We report examples for 3 different GW-injection distances, corresponding to redshifts $\{0.1,0.5,1\}$.

\begin{figure}[H]
    \centering
    \includegraphics[width=.49\linewidth]{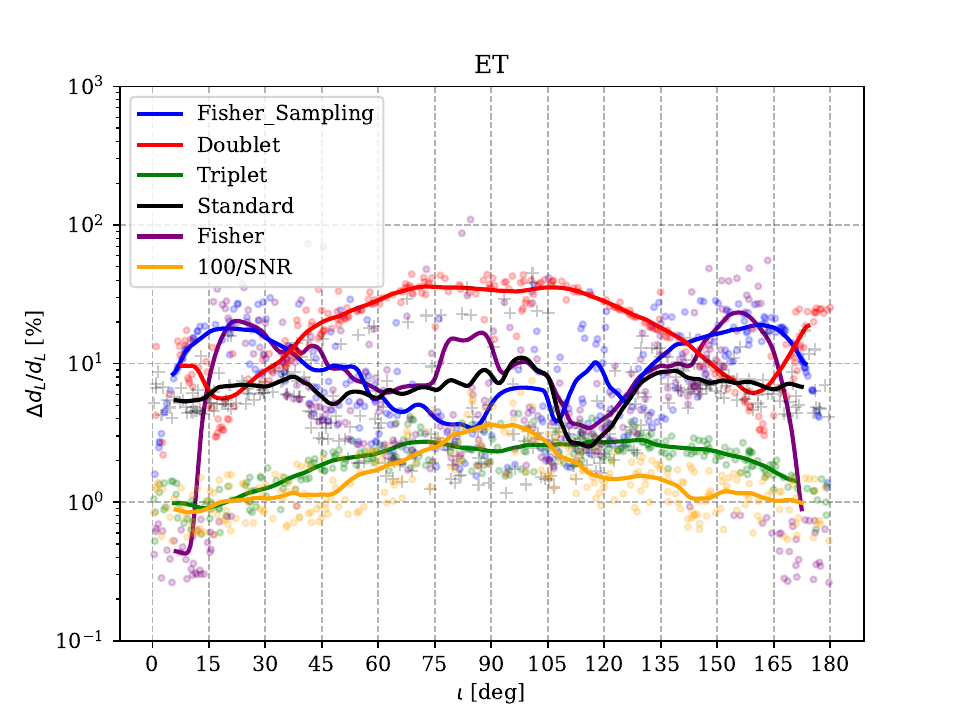}
    \includegraphics[width=.49\linewidth]{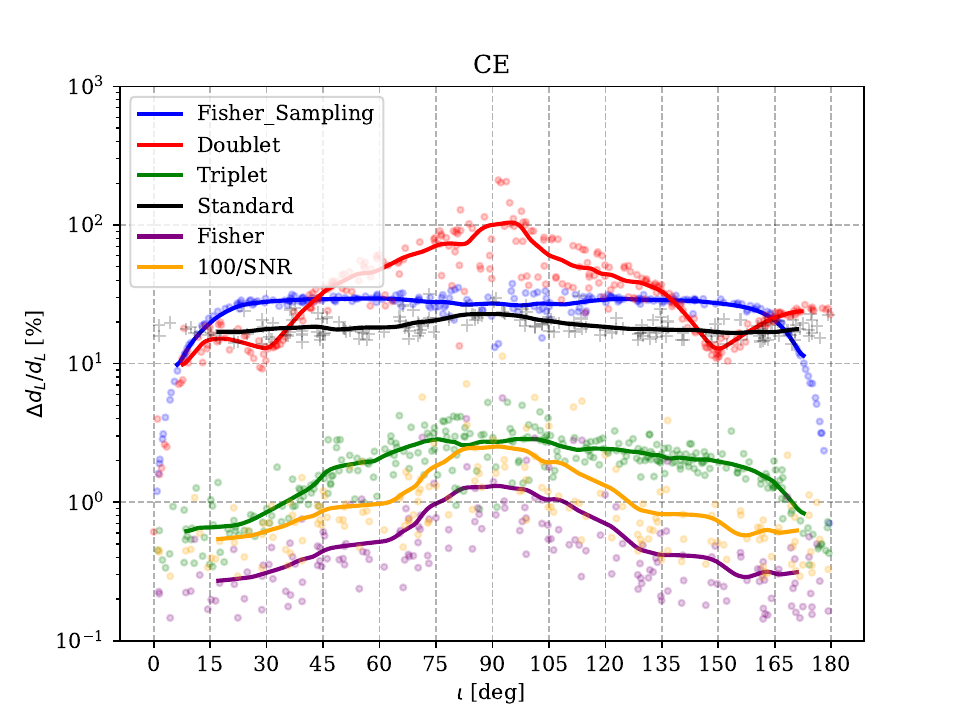}
    \caption{Estimates of $\Delta d_L/d_L$ vs. $\iota$ averaged over 300
      simulations for equal mass, spin-less binary system of fixed total mass $3M_\odot$,
     $d_L\simeq 460$ Mpc corresponding to redshift $z=0.1$
    in the standard $\Lambda CDM$ model with $H_0=70\,km/s/Mpc$ and $\Omega_M=0.3$. Cosmological computation makes use of \textit{astropy} \cite{astropy:2022}.}
    \label{fig:et_ce_01}
\end{figure}

\begin{figure}[H]
    \centering
    \includegraphics[width=.49\linewidth]{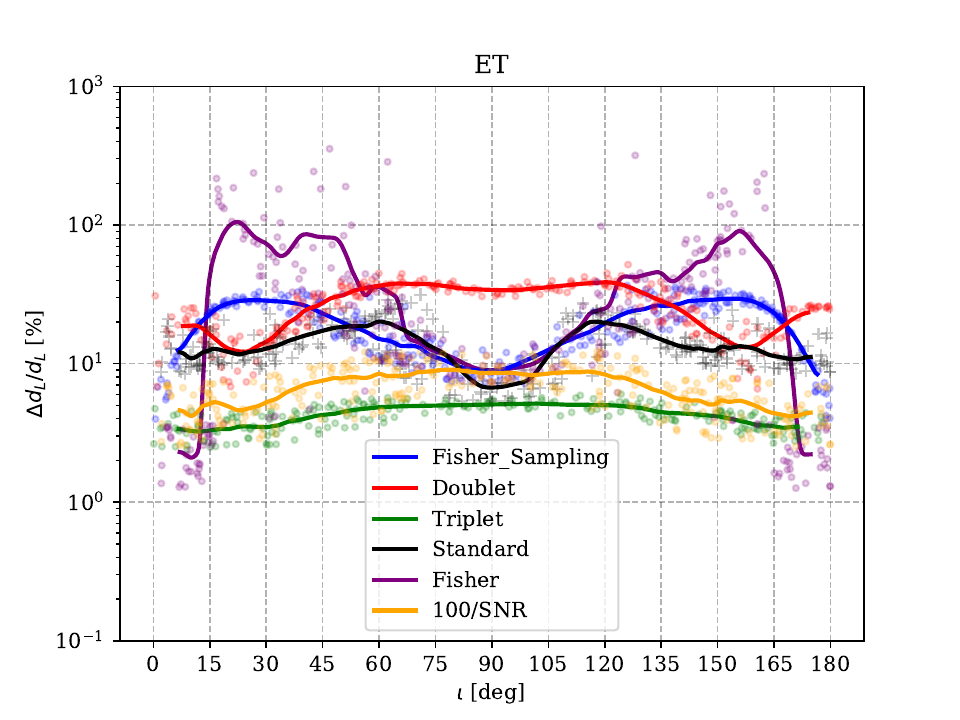}
    \includegraphics[width=.49\linewidth]{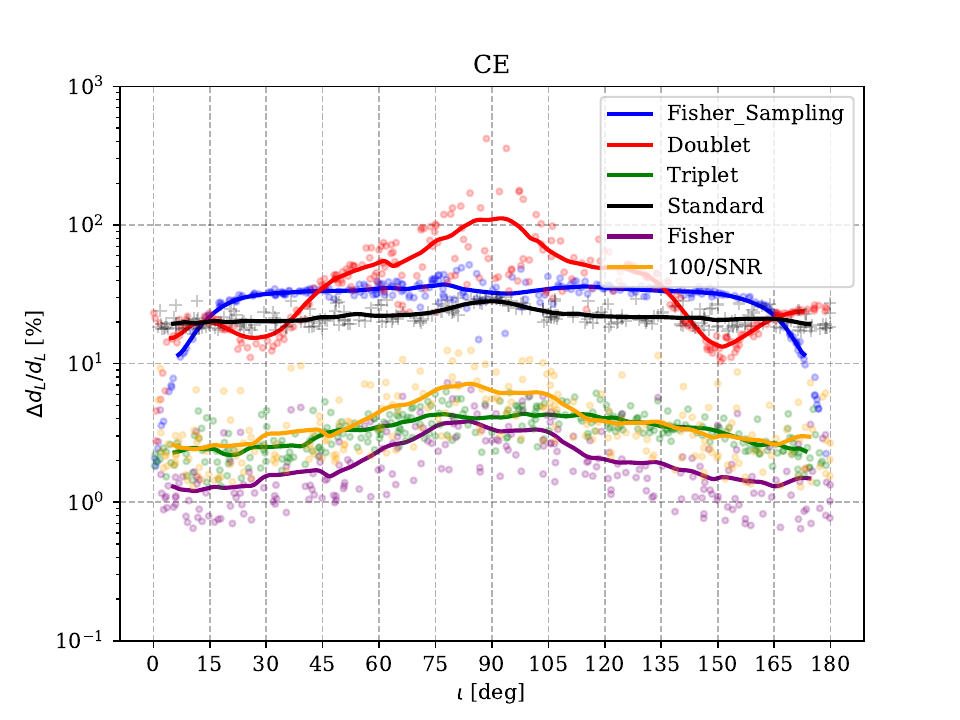}
    \caption{Same as in Fig.~\ref{fig:et_ce_01}, but with redshift $z=0.5$, or $d_L\simeq 2.8$ Gpc.}
    \label{fig:et_ce_05}
\end{figure}

\begin{figure}[H]
    \centering
    \includegraphics[width=.49\linewidth]{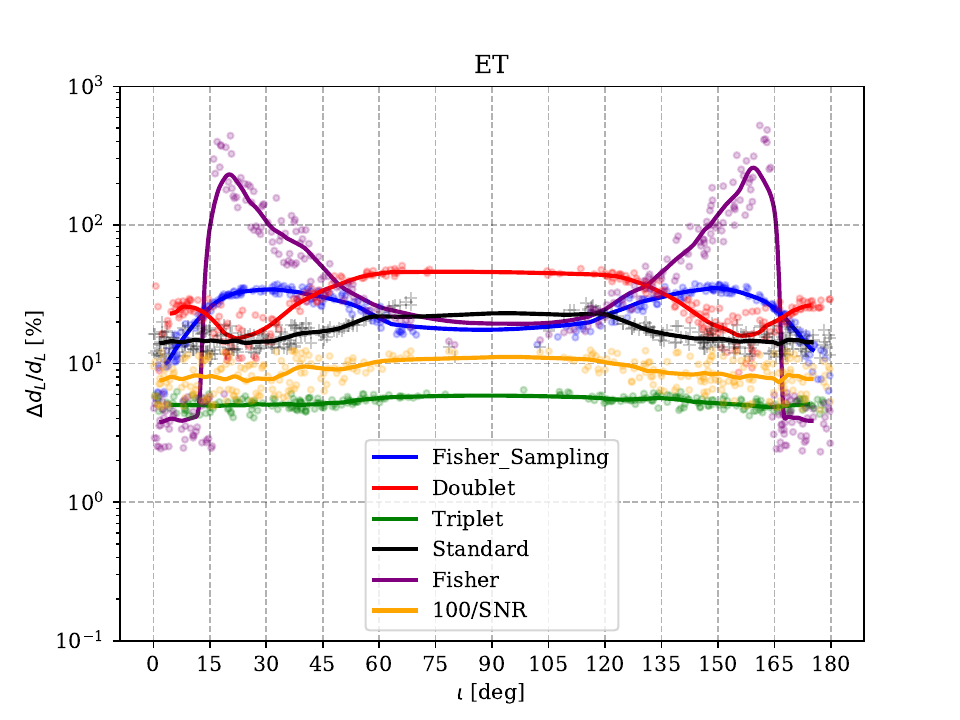}
    \includegraphics[width=.49\linewidth]{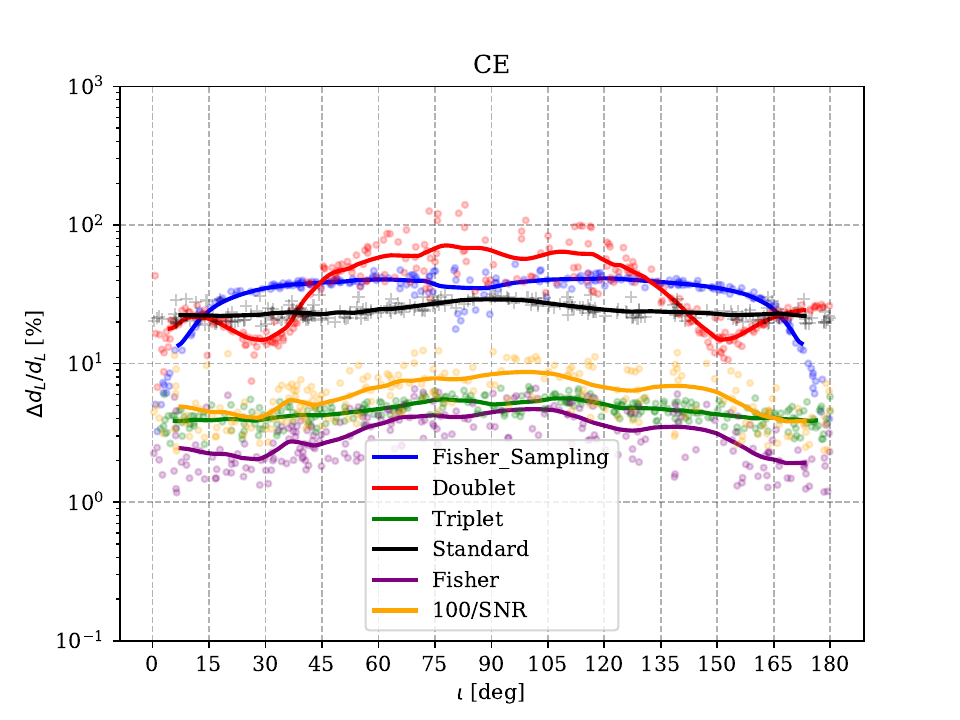}
    \caption{Same as in Fig.~\ref{fig:et_ce_01}, but with redshift $z=1$, or $d_L\simeq 6.6$ Gpc.}
    \label{fig:et_ce_10}
\end{figure}

Figs.\ref{fig:et_ce_01},\ref{fig:et_ce_05},\ref{fig:et_ce_10} report
the relative uncertainties in $d_L$, each figure consisting of two plots:
the one
on the left is for a triangle-shaped detector as ET, the one on
the right for a $L$-shaped one as for CE, with spectral
noise sensitivities as in Fig.~\ref{fig:sns}.

One can see the "turn around" in the Fisher matrix derived
uncertainty of $d_L$, which overall grows for $\iota\to 0,\pi$
until reaching a maximum value and then decreasing.
This is due to the onset of the above mentioned Penrose method, which prevents taking the inverse of the full matrix
when the minimum eigenvalue (in absolute value) goes below a threshold, which in GWDALI has been fixed to a the moderate value of $10^{-4}$ (\verb|rcond| parameter shown in the example code above).
For minimum eigenvalues (in absolute value) lower than the threshold, the Fisher matrix
is effectively treated as lower dimensional.
In the ET case, the Fisher matrix has no vanishing eigenvalue
unless $\iota=0,\pi$, where it gains 1 zero eigenvalue.
The $d_L$ uncertainty grows for $\iota\to 0,\pi$ until 
the Penrose pseudo-inverse method excludes the flat direction
by effectively reducing the dimensionality of the Fisher matrix.

In the ET case one has 1 zero eigenvalue for \emph{any} value of $\iota$, hence the flat direction is always chopped out.

On the other hand the Doublet and Triplet approximations do not
present pathological behaviour, broadly following the full-likelihood result. For completeness, in \ref{app:bias} we report the relative difference 
between the innjected value of $d_L$ and its maximum likelihood value for
methods involving a MonteCarlo sampling of the (exact or approximate) likelihood, showing comparable behaviour of the Doublet and Triplet approximation and
of the exact likelihood.

\section{Conclusions}
\label{sec:concl}

We have presented a novel python package for gravitational wave parameter
uncertainty forecast based on extending the log-likelihood
derivative approximation to higher level than two derivatives.
The main scope is to avoid large, unphysical uncertainty forecast for parameters
when the Fisher matrix is not invertible.

This software is highly flexible as it allows to call the {\tt LAL} waveform generator
and the stochastic sampling is performed via {\tt Bilby}.
It allows to select an arbitrary detector network by defining detector locations,
orientations and topologies.

A crucial step in the calculation of approximate likelihood is derivative
computation, so making the precision of numerical derivatives as high as
possible is crucial to improve the consistency of the code, which we
tested by comparison with results obtained by MonteCarlo sampling of the
exact likelihood.

Natural extensions of this code include 
cosmological applications \cite{Gray:2019ksv,Gray:2021sew}
with assessment of cosmological parameter precision when studying population instead of individual detections.

\appendix

\section{Derivative expansion beyond cubic order: P- and H-Tensors}
\label{app:PH}

The Taylor expansion of the likelihood (\ref{eq:taylor}) has a 4-th order term
specified by the P-Tensor (\ref{eq:beyond}):
\begin{align*}
  \mathcal{P}_{ijklm} & \equiv-\left(\partial_{i}\partial_{j}\partial_{k}\partial_{l}\partial_{m}log\mathcal{L}\right)_{0}\\
  &= -\left[\left\langle \partial_{l}\partial_{m}h|\partial_{i}\partial_{j}\partial_{k}h\right\rangle +(9\ \rm{perms})\right] -\left[\left\langle \partial_{i}h|\partial_{j}\partial_{k}\partial_{l}\partial_{m}h\right\rangle +(4\ {\rm perms})\right]
\end{align*}
whose contribution to the likelihood is
\begin{align*}
  log\mathcal{L}_5 & =-\frac{1}{5!}\sum_{a,\cdots,e}\mathcal{P}_{abcde}\Delta\theta^{abcde}\\
  & =-\sum_{a,\cdots ,e}\left[\frac 1{12}\left\langle \partial_a\partial_bh|\partial_c\partial_d\partial_eh\right\rangle
  +\frac{1}{24}\left\langle \partial_ah|\partial_b\partial_c\partial_d\partial_eh\right\rangle\right] \Delta\theta^{abcde}\,.
\end{align*}
It contains both 3- and 4- derivative terms, only the former contributing to
the Triplet-DALI.

The $H$-Tensor term is
\begin{align*}
  \mathcal{H}_{abcdef} \equiv&-\left(\partial_a\partial_b\partial_c\partial_d\partial_e\partial_flog\mathcal{L}\right)_{0}\\
  &    -\left[\left\langle \partial_a\partial_b\partial_ch|\partial_d\partial_e\partial_fh\right\rangle +(19\ {\rm perms})\right]
  -\left[\left\langle \partial_a\partial_bh|\partial_c\partial_d\partial_e\partial_fh\right\rangle +(14\ {\rm perms})\right]\\
      &-\left[\left\langle \partial_ah|\partial_b\partial_c\partial_d\partial_e\partial_fh\right\rangle +(5\ {\rm perms})\right]\,,
\end{align*}
contributing to the likelihood
\begin{align*}
  log\mathcal{L}_6 & =-\frac 1{6!}\sum_{a,\cdots,f}\mathcal{H}_{abcdef}\Delta\theta^{abcdef}\\
  & =-\sum_{a,\cdots,f}\left[
    \frac{1}{72}\left\langle\partial_a\partial_b\partial_ch|\partial_d\partial_e\partial_fh\right\rangle
    +\frac1{48}\left\langle \partial_a\partial_bh|\partial_c\partial_d\partial_e\partial_fh\right\rangle\right.\\
    &\qquad\qquad \left.+\frac 1{120}\left\langle \partial_ah|\partial_b\partial_c\partial_d\partial_e\partial_fh\right\rangle\right]
  \Delta\theta^{abcdef}\,.
\end{align*}
Collecting the 3-derivative terms from Quarxion, P- and H-tensors we have
the Triplet contribution to the likelihood:
\begin{align}
  \left. \log{\cal L}\right|_{Triplet}=
  & -\frac 16\sum_{a,b,c,d}\left\langle \partial_ah|\partial_b\partial_c\partial_dh\right\rangle\Delta\theta^{abcd}
	-\frac{1}{12}\sum_{a,b,c,d,e}\left\langle \partial_a\partial_bh|\partial_c\partial_d\partial_eh\right\rangle\Delta\theta^{abcde}\\
	&-\frac{1}{72}\sum_{a,\dots ,f}\left\langle \partial_a\partial_b\partial_ch|\partial_d\partial_e\partial_fh\right\rangle \Delta\theta^{abcdef}
\end{align}

\section{Example of parameter recovery}
\label{app:bias}

\begin{figure}[H]
    \centering
    \includegraphics[width=.49\linewidth]{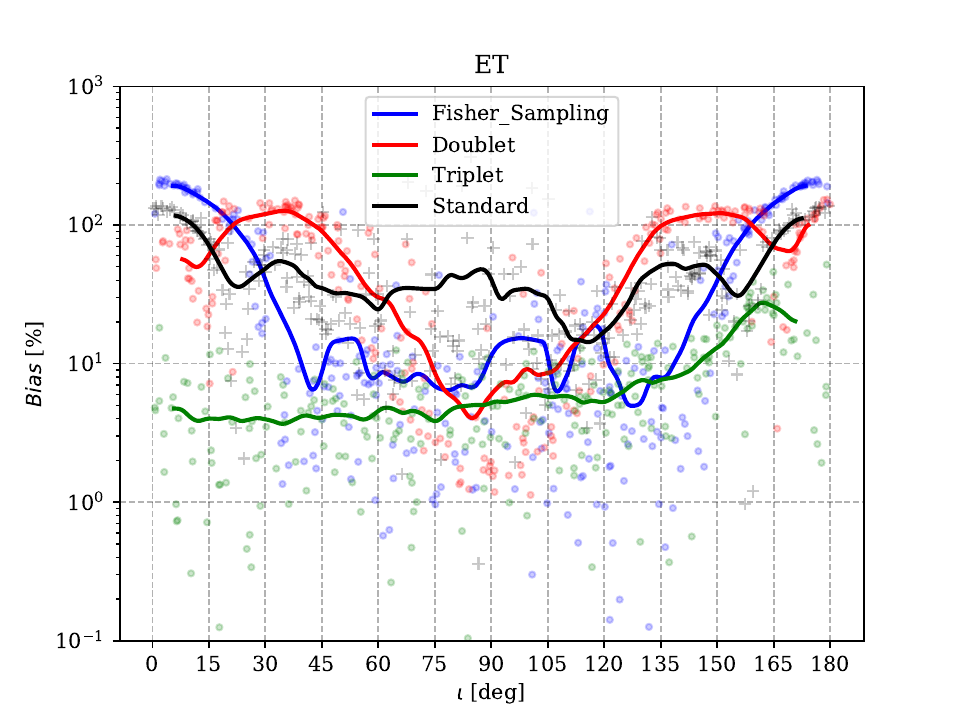}
    \includegraphics[width=.49\linewidth]{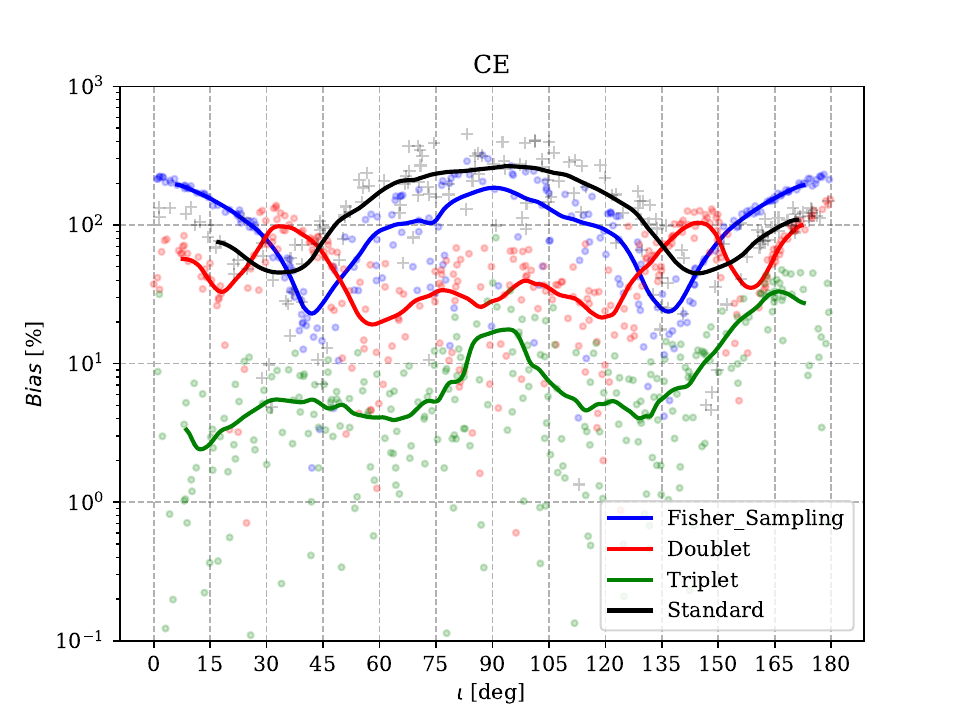}
    \caption{$d_L$ relative bias for the same injections as in Fig.~\ref{fig:et_ce_01}.}
    \label{fig:et_ce_bias01}
\end{figure}

\begin{figure}[H]
    \centering
    \includegraphics[width=.49\linewidth]{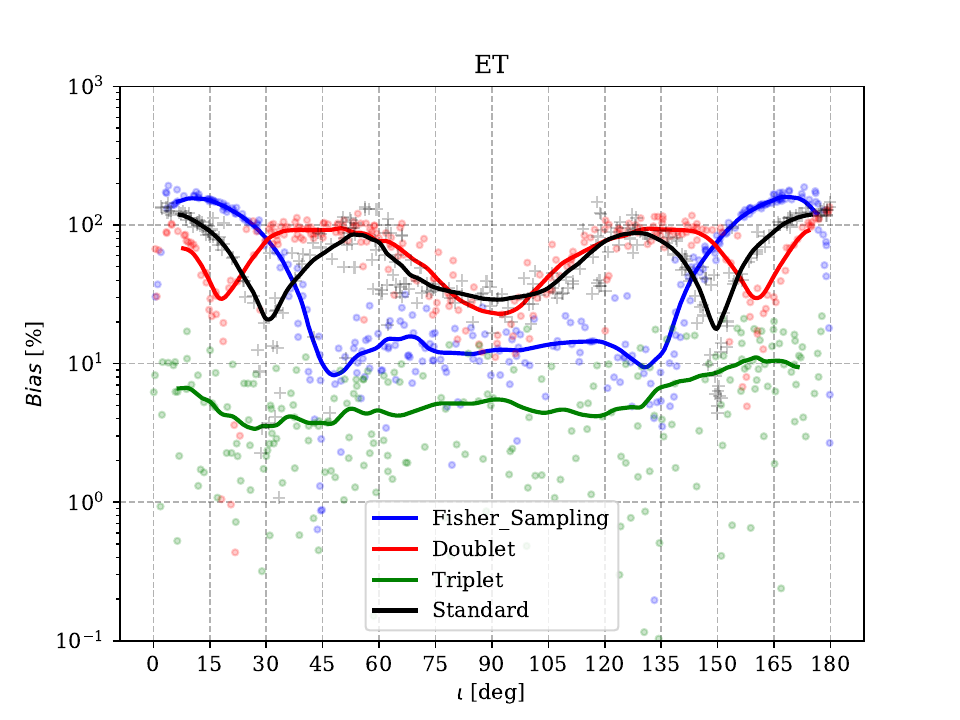}
    \includegraphics[width=.49\linewidth]{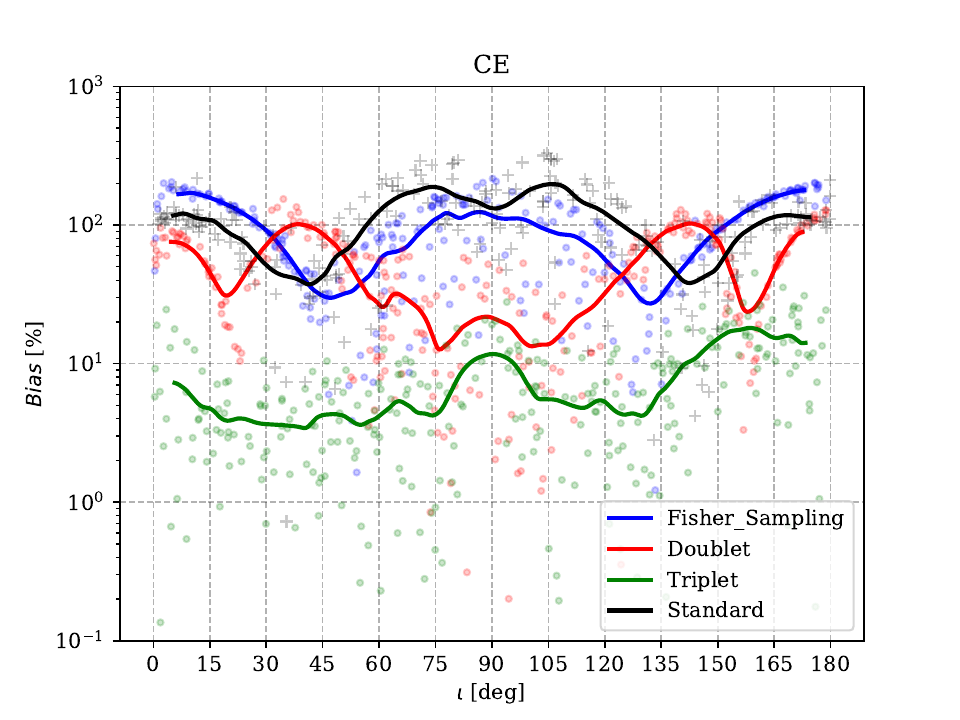}
    \caption{Same as in Fig.~\ref{fig:et_ce_bias01}, but for $z=0.5$ ($d_L\simeq 2.8$Gpc).}
    \label{fig:et_ce_bias05}
\end{figure}

\begin{figure}[H]
    \centering
    \includegraphics[width=.49\linewidth]{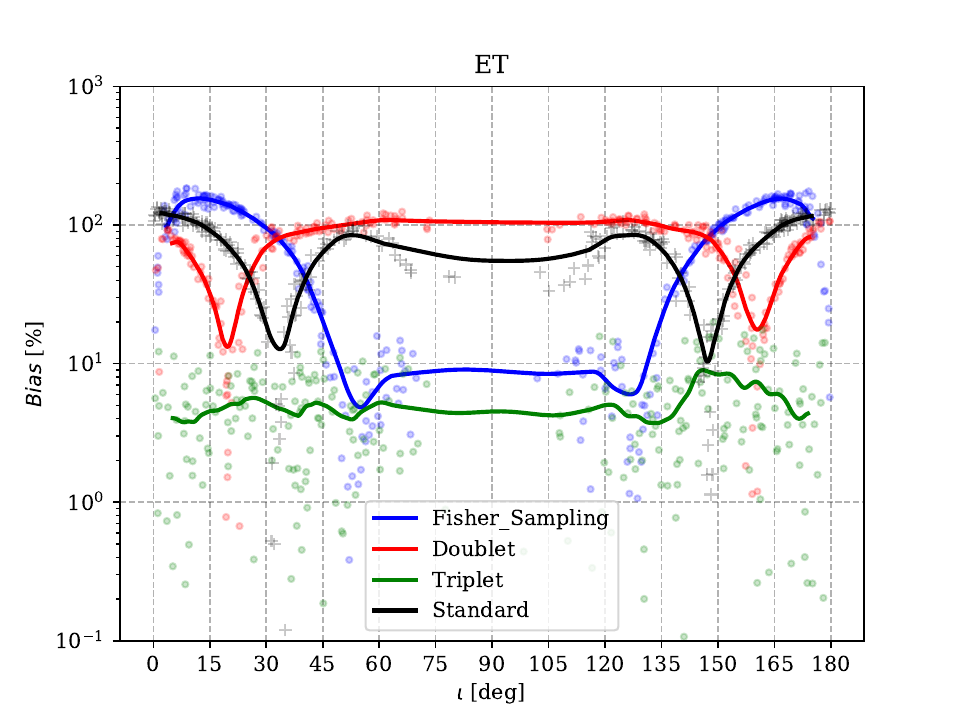}
    \includegraphics[width=.49\linewidth]{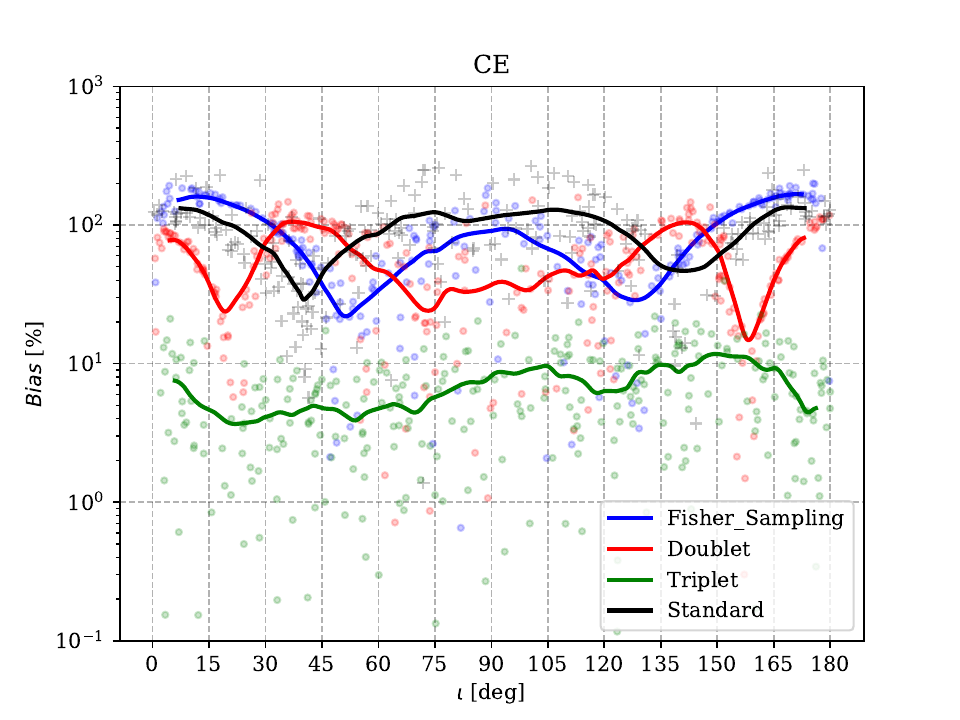}
    \caption{Same as in Fig.~\ref{fig:et_ce_bias01}, but for $z=1$
      ($d_L\simeq 6.6$Gpc).}
    \label{fig:et_ce_bias10}
\end{figure}

For the analysis of subec.~\ref{ssec:exs} we report in Figs.~\ref{fig:et_ce_bias01},\ref{fig:et_ce_bias05},\ref{fig:et_ce_bias10}, the relative bias,
i.e. the relative difference between the injected and maximum likelihood
estimator recovered with different likelihood derivative approximations and
with the exact likelihood, running the {\tt Nestle} \cite{Mukherjee:2005wg} MonteCarlo sampler
on each of them.

\section*{Acknowledgments}
  The authors thank Valerio Marra for useful discussions.
  JMSdS is supported  by the Coordena\c{c}\~ao de Aperfei\c{c}oamento de Pessoal de
  N\'\i vel
  Superior (CAPES)  -- Graduate Research Fellowship/Code 001.
  The work of RS is partially supported by CNPq under grant 310165/2021-0 and
  by FAPESP grants 2021/14335-0 and 2022/06350-2.
  The authors thank the High Performance Computing Center (NPAD) at UFRN for
  providing computational resources that made the present work possible.

\bibliographystyle{elsarticle-num}

\end{document}